\documentclass[pra,aps,twocolumn,superscriptaddress,notitlepage]{revtex4-2}
\usepackage{amssymb,amsfonts,amsmath,amstext,graphicx,hyperref,lipsum,empheq}
\usepackage{float}
\usepackage[normalem]{ulem}
\hypersetup{
	colorlinks=true,
	linkcolor=blue,
	filecolor=magenta,      
	urlcolor=cyan,
}
\usepackage{amsthm}
\usepackage{outlines}
\usepackage{subfigure}

\usepackage[dvipsnames]{xcolor}
\usepackage{hyperref}

\newcommand{\beq}[0]{\begin{equation}}
\newcommand{\eeq}[0]{\end{equation}}

\def\be{\begin{equation}}
\def\ee{\end{equation}}
\def\bea{\begin{eqnarray}}
\def\eea{\end{eqnarray}}
\newcommand{\ba}{\begin{eqnarray}}
\newcommand{\ea}{\end{eqnarray}}
\newcommand{\kB}{k_\mathrm{B}}

\def\tr{\mbox{tr}}

\def\bra#1{\langle{#1}|}
\def\ket#1{|{#1}\rangle}

\def\Bra#1{\left\langle#1\right|}
\def\Ket#1{\left|#1\right \rangle}
\def\BraVert{\egroup\,\mid\,\bgroup}

\definecolor{myblue}{rgb}{.8, .8, 1}

\usepackage{multirow}	

\begin{document}

\title{Quadratic enhancement in the reliability of collective quantum engines}
\author{Noufal Jaseem}
\affiliation{Department of Physical Sciences, Indian Institute of Science Education and Research Berhampur, Berhampur 760010, India.}

\author{Sai Vinjanampathy}
\affiliation{Department of Physics, Indian Institute of Technology-Bombay, Powai, Mumbai 400076, India.}
\affiliation{Centre for Quantum Technologies, National University of Singapore, 3 Science Drive 2, 117543 Singapore, Singapore.}

\author{Victor Mukherjee }
\affiliation{Department of Physical Sciences, Indian Institute of Science Education and Research Berhampur, Berhampur 760010, India.}

\begin{abstract}
   { We study fluctuations in many-body quantum heat engines operating in the presence of collective system-bath interactions.  We show that collective effects in open quantum systems can be harnessed to develop highly consistent many-body quantum engines. We consider quantum Otto engines, modeled by $n$ spins collectively coupled to thermal baths.  Our results show that collective effects can significantly reduce the fluctuations in the output work, quantified by high reliability ($r$)  and low thermodynamic uncertainty.  In contrast to independent engines, we demonstrate a quadratic enhancement of the reliability $r$ for their collective counterparts. We extend our analysis to the case of interacting spin models commonly studied in many-body physics, such as the Lipkin-Meshkov-Glick (LMG) model, thereby broadening the regime of applicability of collective effects in quantum thermal machines significantly. This paves the way forward for realistic collective quantum thermal machines in many body systems.} 
   
\end{abstract}

\maketitle

\date{\today}


\section{Introduction}
Modeling and control of quantum systems are crucial for the development of quantum technologies that exhibit an advantage over their classical analogs \cite{kosloff14quantum, klimovsky15thermodynamics, vinjanampathy16quantum, goold16the, bhattacharjee21quantum, mukherjee21many, myers22quantum} such as computing, sensors and quantum thermal machines~\cite{PhysRevLett.86.5188,kielpinski2002architecture,monroe2014large,PhysRevA.46.R6797,PhysRevD.23.1693,giovannetti2004quantum,PhysRevA.33.4033,PhysRevA.57.4736,PhysRevA.54.R4649,demkowiczchapter,dowling2008quantum,jaseem2018quantum,jaseem2017two,kosloff14quantum, klimovsky15thermodynamics, vinjanampathy16quantum, goold16the, bhattacharjee21quantum, mukherjee21many,myers22quantum,PhysRevLett.125.240602,PhysRevLett.127.190604}. Each of these technologies is assessed by a performance benchmark, for instance, gate complexity for computing~\cite{PhysRevA.82.042305,PhysRevA.82.042319} and variance of an unbiased estimator for sensors~\cite{PhysRevA.46.R6797,giovannetti2004quantum,PhysRevA.33.4033,PhysRevA.57.4736,PhysRevA.54.R4649,demkowiczchapter,dowling2008quantum,jaseem2018quantum,jaseem2017two}. The performance of quantum thermal machines, on the other hand, has been evaluated through three key indicators, namely average efficiency, average output power, and more recently reliability ($r$), quantified through the ratio of the average work
to its root-mean-square deviation \cite{esposito09nonequilibrium}. In this regard, various avenues related to quantum correlations in small quantum systems, such as non-Markovianity~\cite{breuer02, mukherjee20anti, das20quantum, wiedmann21non,modi}, optimal paths~\cite{suri,erdman19maximum, vasco21maximum, erdman22identifying, khait22optimal} and non-thermal baths \cite{rossnagel14nanoscale, niedenzu18quantum} have been shown to enhance the performance of quantum machines. Following such theoretical investigations, several experimental realizations \cite{rossnage16a,jan17squeezed,mslennikov19quantum,klatzow19experimental,pal20experimental} of microscopic thermal machines have led to significant advancements in the fields of quantum thermodynamics and quantum technologies. 

Many-body quantum systems offer unique opportunities to engineer quantum technologies operating in the presence of non-trivial many-body effects, such as phase transitions \cite{campisi16the, revathy20universal, fogarty20a}, localization-delocalization transitions \cite{halpern19quantum, chiaracane20quasiperiodic} and cooperative effects arising due to collective interactions with the environment \cite{niedenzu18cooperative}. Interestingly, collective effects have been shown to be highly beneficial in many quantum technologies \cite{carlos, angsar19collectively,  latune20collective, watanabe20quantum,seeding, souza22collective, kamimura22quantum}, and have been used to enhance efficiency \cite{klimovsky19cooperative} and work output \cite{niedenzu18cooperative, kloc19collective,  latune20collective} in quantum heat engines, and perform high-precision quantum thermometry \cite{latune20collective}. What remains an open question is if quantum correlations arising due to the environment can improve the consistency of quantum thermal machines. In this manuscript, we answer this crucial question in the affirmative and show that cooperative quantum effects can stabilize normalized fluctuations of a many-body engine better than their non-cooperative analogs. In general, reliability $r$ in machines composed of  $n$ subsystems scales as $\sqrt{n}$ \cite{reif09}. This leads to increased fluctuations in the microscopic regime, in machines comprising a finite number of particles. In contrast, here we show that collective effects can result in the reliability $r$ scaling as $n$, thus making collective quantum heat engines highly reliable, and paving a possible way for developing realistic and reliable quantum technologies. We begin by reviewing a generic model of a quantum engine and discuss the metrics for quantifying the performance of such an engine. For a quantum engine based on a generic many-body working medium (WM), we show that a broad class of steady-states results in a $r \sim n$ scaling of the reliability, thus emphasizing the wide applicability of our results. We then exemplify our generic theory using specific examples of non-interacting as well as interacting multi-spin WMs. Finally, we summarize our main results. 

\section{Reliable Collective Quantum Thermal Machines}
The performance of quantum thermal machines is assessed based on a variety of performance measures that examine the quality, quantity, and time in which a thermodynamic task is performed. In the case of quantum engines, two well-known metrics are output work and the efficiency of the engine. The third metric of performance is reliability, which measures the quality of output work considering it to be a fluctuating quantity. Reliability is defined as $r=\langle W\rangle/\Delta W$, where $\langle W^p\rangle:=\int_{-\infty}^{\infty} dW P(W)W^p$ is the moment of the work distribution function and $\Delta W:=\sqrt{\langle W^2\rangle-\langle W\rangle^2}$ is the standard deviation. Reliability hence measures the consistency of the output work.
 Reliability was investigated in a series of studies~\cite{juzar2021monitoring,sacchi2021multilevel,Sacchi2021boson,agarwalla2018assessing,ptaszynski2018coherence,guarnieri2019thermodynamics,cangemi2020violation,kalaee2021violating}. For instance, in~\cite{juzar2021monitoring} the impacts of diagnostic schemes to determine the performance of quantum Otto heat engines on different figures of merit such as reliability were explored. In~\cite{sacchi2021multilevel}, a two-qudit SWAP engine was studied and it was shown that for fermionic SWAP engines the reliability is bound by $r^2\leq \langle \Sigma\rangle/(2-\langle \Sigma\rangle)$, where $\langle \Sigma\rangle$ is the entropy production. The bosonic SWAP engines, on the other hand,  obeys $r^2\leq \langle \Sigma\rangle/(2+\langle \Sigma\rangle)$~\cite{Sacchi2021boson}.
\begin{figure*}
\centering
    \subfigure[]{\includegraphics[width=0.3\textwidth]{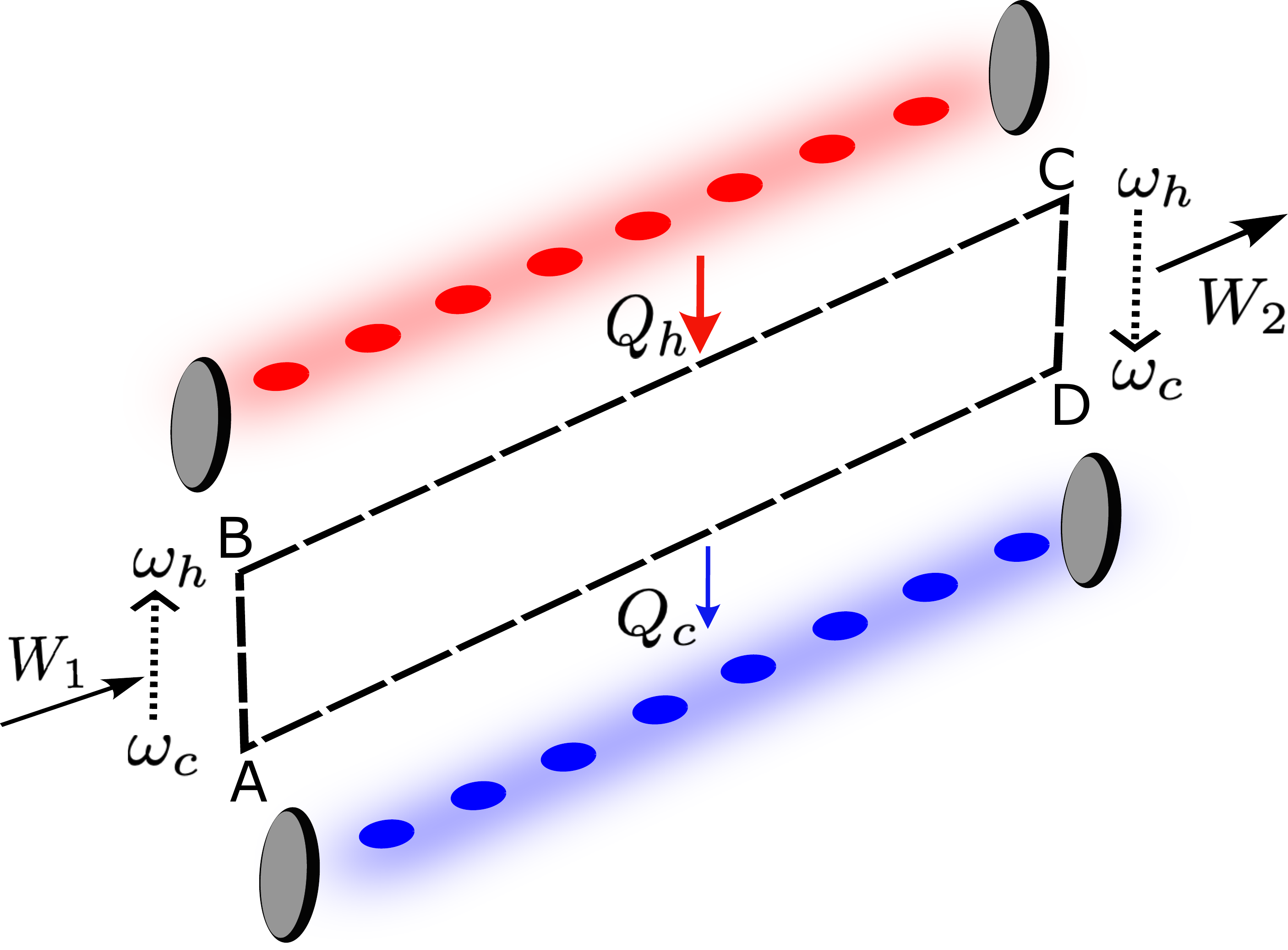} \label{fig:otto_schematic}}\hfill
    \subfigure[]{\includegraphics[width=0.32\textwidth]{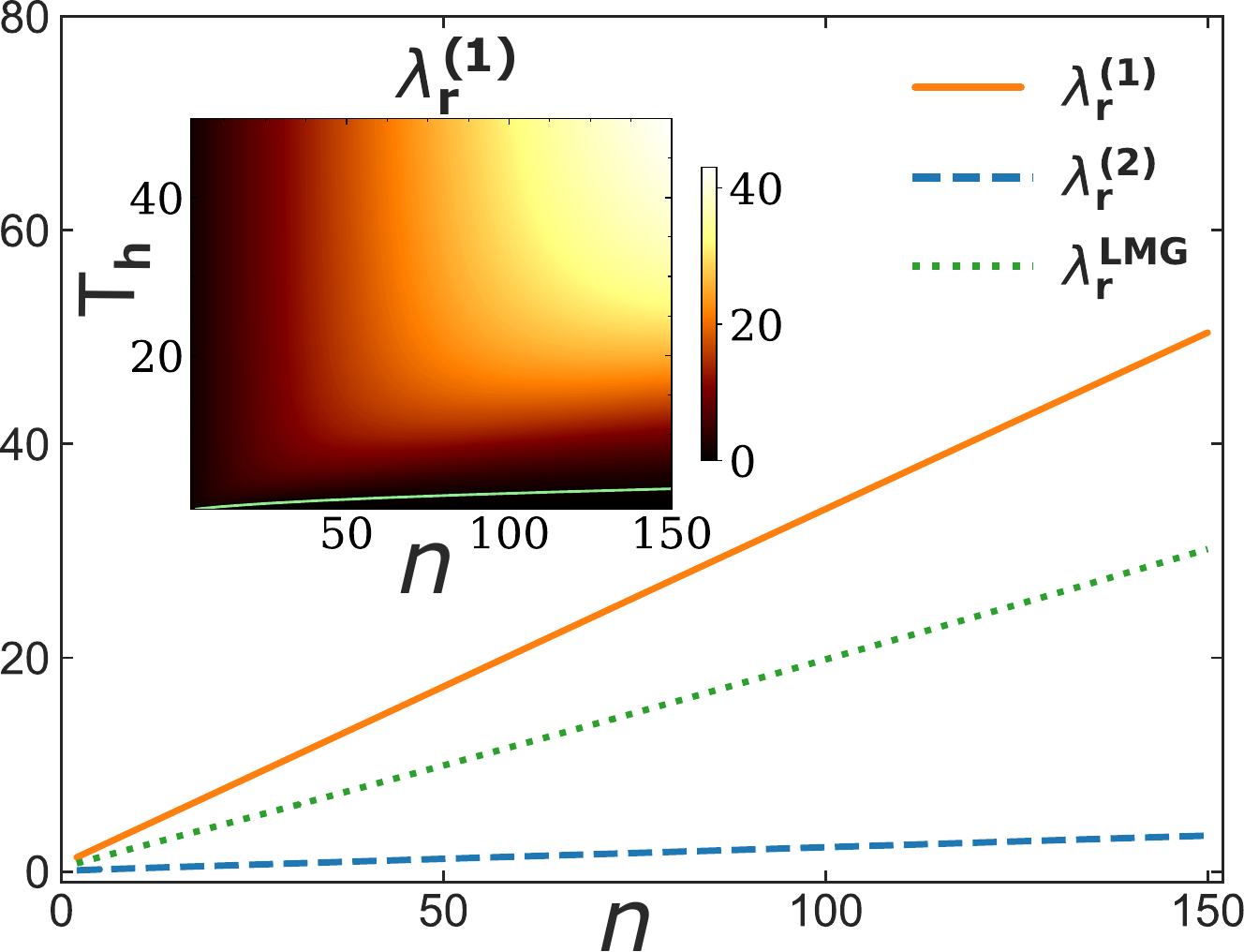} \label{fig:r_ratio_den}} \hfill 
    \subfigure[]{\includegraphics[width=0.32\textwidth]{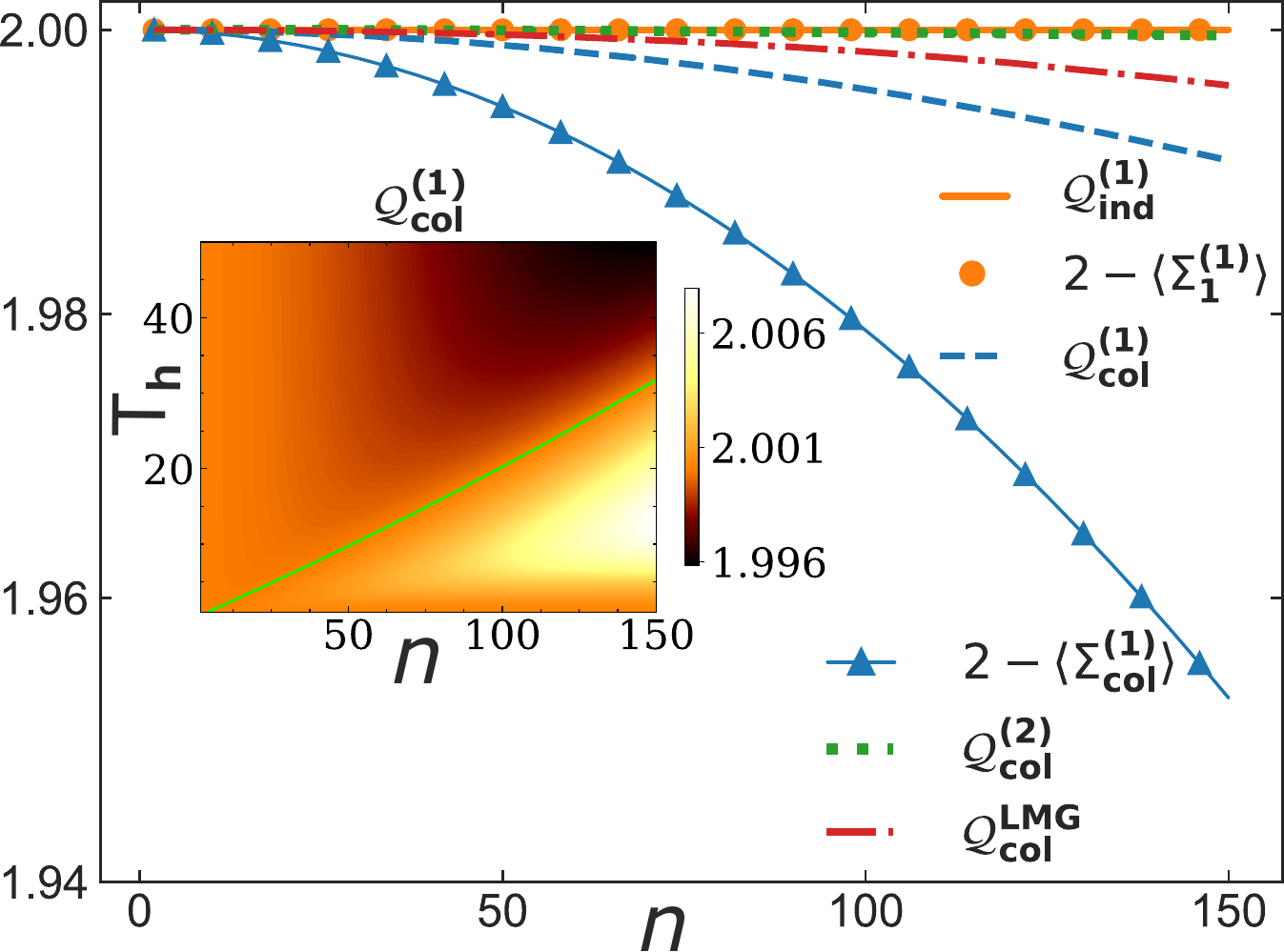}  \label{fig:tur}} 
    \caption{ (a) Schematic diagram of an Otto engine with an $n$-spin WM interacting collectively with thermal baths. The energy gap of the engine system is varied along the paths {\bf A $\to$ B} and {\bf C $\to$ D}, while the system interacts collectively with baths along paths {\bf B $\to$ C} and {\bf D $\to$ A}.
    (b) The ratio of reliability of the collective engine to that of the independent engine, $\lambda_r$, is plotted against $n$ for the different Hamiltonian given in Eqs.~\eqref{eq:HLMG} and \eqref{eq:Hint}. Inset: $\lambda_r^{(1)}$ is plotted as a function of $n$ and $T_{\rm h}$, for $x = 1$. The advantage of collective QHEs over independent ones becomes more pronounced for large $T_{\rm h}$ and $n$. The green line is the contour line for $\lambda_r =1$. (c) The left and right-hand sides of the TUR, in Eq.(\ref{eq:tur}), are plotted as a function of the number of spins $n$ for the collective and independent cases. For the independent case, $\mathcal{Q}_{\rm ind}\geq 2-\langle \Sigma_1^{(1)}\rangle$, where $\langle \Sigma_1^{(1)}\rangle$ is the single qubit entropy production (see Appendix~\ref{App_sec_tur}). Inset: $\mathcal{Q}_{\rm col}^{(1)}$ is plotted as a function of $n$ and $T_{\rm h}$. Low $\mathcal{Q}_{\rm col}^{(1)}$ for large $n$ and $T_{\rm h}$ implies the collective engine is most beneficial in this regime. The green line is the contour line for $\mathcal{Q}_{\rm col}^{(1)} =2$. For both (a) and (b), the parameter values are fixed as $\omega_h=0.5$, $\omega_c=0.1$, $\gamma=0.7$, and $\Delta =\beta_c\omega_c-\beta_h\omega_h= 0.005$. For the main plots, $\beta_h\to 0$ and for the insets, both $T_{\rm h}$ and $T_{\rm c}$ are varied so that $\Delta$ is kept constant. Superscripts label $x=1,2$ or LMG.}
\end{figure*}
The different moments of the work distribution function are related to each other via the fluctuation-dissipation theorem, which can be formalized in terms of a thermodynamic uncertainty relation (TUR)~\cite{ito2019universal,kalaee2021violating,horowitz2020thermodynamic,sacchi2021multilevel,Sacchi2021boson}. A natural question that arises is whether the reliability of the engine can be enhanced by collective interactions. Consider $n$ non-interacting quantum thermal machines. It is easy to see that the work and variance scale extensively with the size of the system \cite{reif09}. This implies that the reliability scales as $\mathcal{O}(\sqrt{n})$ which can be understood as the lower bound imposed by non-interacting systems \cite{reif09}. In contrast to this, here we consider a generic many-body quantum engine collectively coupled to thermal baths, as depicted in Fig.~\ref{fig:otto_schematic}, operating near the Carnot efficiency, $\eta_{\rm C}$. It was previously established that for non-interacting collective quantum heat engines the mean output work scales as the collective heat capacity $C^{\rm col}$ in this case~\cite{latune20collective}, see the Appendix \ref{App_sec_work}. In this manuscript, we establish that, for a more generic setup, the variance of the output work also scales with $C^{\rm col}$, which implies that so does $r^2$. Now, to understand the collective advantage in the reliability, let the generic non-equilibrium steady state of a $n$-body quantum engine operating near Carnot efficiency be written as  
$\rho(\beta)=\exp(-\beta n \mathcal{A})/\tr\left[\exp(-\beta n \mathcal{A})\right]$
where $\beta$ is the inverse temperature of a bath and $\mathcal{A}$ is an operator in the collective basis which does not scale with system size. In this case, as we show below for  appropriate  $\mathcal{A}$, $C^{\rm col}\propto {n}^2$ and hence the reliability scales as $n$ (see Appendices \ref{App_sec_var}, \ref{App_sec_scaling}, and \ref{App_sec_LMG} for details). This is a quadratic enhancement in the reliability of generic collective quantum thermal machines. 

\section{Many Body Otto Engine and Statistics} 
To illustrate and formalize our general result, we consider an Otto cycle of a many-body collective heat engine. The Otto cycle, described in  Fig. \ref{fig:otto_schematic} consists of four strokes, namely the following. (i) ({\bf A $\to$ B}): We assume the WM  starts from the steady state with respect to a thermal bath at temperature $T_{\rm c}$. The frequency $\omega$ of the WM is increased from $\omega = \omega_{\rm c}$ to $\omega = \omega_{\rm h}>\omega_{\rm c}$ during this unitary stroke, thereby performing $W_1$ amount of work on the system. (ii) ({\bf B $\to$ C}): The WM is coupled to a hot thermal bath during this non-unitary stroke, while the frequency is kept constant at $\omega = \omega_{\rm h}$; $Q_{h}$ is the amount of heat flowing from the hot bath to the WM, such that the WM reaches the steady state at the end of the stroke. (iii) ({\bf C $\to$ D}): The WM is decoupled from the bath after which the frequency is decreased from $\omega = \omega_{\rm h}$ to $\omega = \omega_{\rm c}$ during this unitary stroke; $W_2$ is the amount of work performed by the WM. (iv) ({\bf D $\to$ A}): The WM is coupled to a cold thermal bath during this non-unitary stroke, while its frequency is kept constant at $\omega = \omega_{\rm c}$; $Q_{c}$ is the amount of heat flowing from the WM to the cold bath, such that the WM reaches the steady state at the end of the stroke, thereby completing the cycle. 

The WM consists of multiple interacting or non-interacting identical spin $s$ particles, such as the LMG model~\cite{lipkin1965validity,Caneva_LMG,larson2010circuit,PhysRevC.104.024305}
\begin{equation} H= \omega(t) \left[ \frac{1}{n}(\mathcal{J}_x^2+\mathcal{J}_y^2)+\gamma \mathcal{J}_z \right], \label{eq:HLMG} \end{equation}
or $x$-body inter-particle interactions, 
\begin{equation} H=n\omega(t) (\mathcal{J}_z/n)^x~~~~\text{for~$x = 1,2,3,...$}.\label{eq:Hint}\end{equation} 
The collective angular momentum is defined as $\mathcal{J}_i := \sum_{k=1}^n J^k_{i}$, where $J_i^k$ is the spin operator for the $k$th spin, along the direction $i = x, y,z$. The spins collectively exchange energy with the thermal baths during the non-unitary strokes. We start by focusing on the $x=1$ case, with $H= \omega(t) \mathcal{J}_z$ and the WM-bath interaction Hamiltonian $H_{\rm int}= \gamma \mathcal{J}_x \otimes \mathcal{B}_{\rm v}$ (${\rm v} = h, c$), where, $\gamma$ denotes the interaction strength, while $\mathcal{B}_c$ ($\mathcal{B}_h$) is the bath operator for the cold (hot) bath. This collective system-bath interaction may result in non-zero off-diagonal terms in the steady-state density matrix $\rho^{\rm ss}$ of the WM at the end of non-unitary strokes in the local spin basis, signifying non-trivial steady-state quantum coherence in the many-body system. The dynamics of the WM are described by the master equation~\cite{niedenzu18cooperative, latune2019thermodynamics, latune20collective},
\begin{eqnarray}
\frac{d\rho}{dt}=-i[H,\rho]+ \sum_{{\rm v}=c,h}\alpha_{\rm v}\mathcal{L}_{\rm v}\rho.
\label{eqmaster}
\end{eqnarray}
Here, $\mathcal{L}_{\rm v}\rho=\Gamma_{\rm v}(\omega_{\rm v})\mathcal{D}(\mathcal{J}^+)\rho+\Gamma_{\rm v}(-\omega_{\rm v}) \mathcal{D}(\mathcal{J}^-)\rho $, $\mathcal{D}[O]\rho =( 2 O^\dagger \rho O- O O^\dagger \rho - \rho O O^\dagger)/2$, $\mathcal{\mathcal{J}}^\pm := \mathcal{\mathcal{J}}_x \pm i \mathcal{\mathcal{J}}_y$ are the collective ladder operators of the spin system and $\Gamma_{\rm v}(\nu)$ denotes the spectral function of the thermal baths at frequency $\nu$, where we consider the Kubo-Martin-Schwinger condition $\Gamma_{\rm v}(-\nu) = \exp\left(-\beta_{\rm v} {\nu} \right)\Gamma_{\rm v}(\nu)$ \cite{breuer02}. We take the physical constants $\hbar=\kB=1$. The constant $\alpha_{\rm v}$ is set to one if the system is coupled to the respective thermal bath, otherwise it is zero.

We consider the system to interact collectively with the baths, such that the system steady states at {\bf A} and {\bf C}  for the aforementioned dissipators (cf. Eq. \eqref{eqmaster}) can be written in the collective basis $|j,m\rangle_i$ as  \cite{latune2019thermodynamics}
\begin{eqnarray}\label{eq:rhoss}
\rho^{ss}(\beta,\omega)&=&\sum_{j=j_0}^{ns}\sum_{i=1}^{l_{j}}P_{j,i}\; \rho_{j,i}^\text{th}(\beta,\omega),
\end{eqnarray}
  where $\rho_{j,i}^\text{th}(\beta,\omega) = \frac{1}{Z_j} \sum_{m=-j}^j e^{-\beta m\omega} |j,m\rangle_i { }_i\langle j,m|$ with $Z_j=\sum_{m=-j}^j e^{-\beta m\omega}$. Here $|j,m\rangle_i$ are the common eigenvectors of $\mathcal{J}_z$ and $\mathcal{J}^2=\mathcal{J}_x^2+\mathcal{J}_y^2+\mathcal{J}_z^2$; $-j\leq m\leq j$, $j\in [j_0; ns]$, and $j_0 = 0$ for $s \geq 1$, while for $s = 1/2$, we have $j_0=1/2$ for $n$ odd and $j_0 = 0$ for $n$ even. The index $i\in [1; l_j]$, where $l_j$ denotes the multiplicity of the eigenspaces associated to the eigenvalue $j(j+1)$ of the $\mathcal{J}^2$ operator. A detailed discussion regarding the collective angular momentum operators and the explicit form of $l_j$ are given in ~\cite{mandel1995optical} and summarized in the Appendix~\ref{App_sec_scaling}  $P_{j,i}=\sum_{m=-j}^j\; { }_{i}\langle j,m|\rho_0|j,m\rangle_i$, with $\rho_0$ being the initial state of the WM.

The dependence of $\rho^{ss}$ on the initial state through $P_{j,i}$ [cf. Eq. \eqref{eq:rhoss}] implies that a choice of $P_{j,i}$ can influence the steady-state significantly. We have $l_{j = ns} = 1$, such that for $P_{j=ns}=1$ the steady state becomes 
\begin{equation}\label{eq:dick_ss}
     \rho^{ss}(\beta,\omega)= \sum_{m=-ns}^{ns} \frac{e^{-\beta m\omega}}{Z_{ns}} |ns,m\rangle\langle ns,m|,
\end{equation}
which is a diagonal state in the symmetrical Dicke subspace known to produce a collective advantage \cite{latune20collective}. This is in contrast to the independent coupling between the particles of the WM and the bath, in which each spin thermalizes independently, such that the WM approaches a direct product steady state at the end of a non-unitary stroke \cite{breuer02, niedenzu18cooperative, latune2019thermodynamics, latune20collective}. 

We note that if the steady state is restricted to a spin $j$ subspace, then  $C^{\rm col}/(\beta\omega)^2 \sim j(j+1)$ for $\beta \to 0$. Consequently, even though collective effects remain present for all values of $j$ following the scaling given above, however,  as we discuss below, these effects become most prominent when we restrict ourselves to the $j = ns$ subspace.

\subsection{Statistics of a many-body collective quantum heat engine} 
Since heat and work are fluctuating quantities, a probability distribution may be associated with obtaining a certain amount of work and heat in an engine cycle. This probability distribution can be found using the two-point measurement scheme, where the system is measured at the beginning and at the end of each stroke. Let the observed system states at {\bf A}, {\bf B}, {\bf C}, and {\bf D} be $\ket{\frac{n}{2}, i} \equiv \ket{i}$, $\ket{j}$, $\ket{k}$, and $\ket{l}$ respectively. Then the joint probability $P(W,Q_h)$ of having certain values of net work $W=W_2+W_1$ and heat $Q_h$ during one cycle of the Otto engine can be readily calculated as~\cite{denzler2020thesis,PhysRevE.103.L060103}
\begin{eqnarray} 
P(W,Q_h) &=& \sum_{i,j,k,l} \delta[W_2-(\epsilon_l^{0}-\epsilon_k^{\tau})]\; \delta[Q_h-(\epsilon_k^\tau-\epsilon_j^{\tau})]\nonumber\\
& & \; \delta[W_1-(\epsilon_j^\tau-\epsilon_i^{0})]\;  p_{k\to l}^{CD}\; p_{j\to k}^{BC}\; p_{i\to j}^{AB}\; p_{i}^A .
\end{eqnarray}
The initial occupation probability of the state $\ket{i}$, $p_i^A=e^{-\beta_c\epsilon_i^{0}}/Z_{(0)}$, as the system is in a steady state with inverse temperature $\beta_c$ at A. The adiabatic transitions give $p_{i\to j}^{AB}=\delta_{i,j}$ and $p_{k\to l}^{CD}=\delta_{k,l}$, and the transition probability $p_{j\to k}^{BC}=e^{-\beta_h\epsilon_k^{\tau}}/Z_{(\tau)}$ since the system reaches a steady state with inverse temperature $\beta_h$ at {\bf C}.  Here, $\epsilon_m^{t=0,\tau}$ are the respective energy eigenvalues for the state $\ket{m}$ at the beginning (time $t=0$) and at the end ($t=\tau$) of the unitary stroke {\bf A} $\rightarrow$ {\bf B}, while the $Z_{(t=0,\tau)}$ are  the corresponding partition functions. The associated characteristic function $\chi(\gamma_1,\gamma_2)$ is given by~\cite{esposito09nonequilibrium,denzler2020thesis,PhysRevE.103.L060103} $\chi(\gamma_1,\gamma_2)=\iint_{-\infty}^\infty dW dQ_h\; e^{i\gamma_1 W} e^{i\gamma_2 Q_h} P(W,Q_h)$, which can then be used to compute the various moments of the heat and work distribution function as $\langle W^r Q_h^s\rangle = \frac{\partial^r}{\partial (i\gamma_1)^r}\frac{\partial^s}{\partial (i\gamma_2)^s} \chi(\gamma_1,\gamma_2)\Big\arrowvert_{\gamma_{1,2}=0}$.

\subsection{$n$ qubit collective engine} Next for simplicity we focus on $s = 1/2$ (see Appendix \ref{App_sec_nspin} for generic $s$) $n$ qubit engine with system Hamiltonian  $H=\omega(t)\sum_{i=1}^{n} \sigma_z^i/2$, which interacts collectively with the thermal baths. We consider the regime close to the Carnot efficiency bound, where the work output approaches zero and hence where work fluctuations can be expected to be prominent. To quantify the reliability advantage of a collective quantum heat engine, we define the ratio of reliability of the collective engine to that of the independent engine: $\lambda_r=r^2_{\rm col}/r^2_{\rm ind}$. In the limit $T_{\rm h} \to \infty$ and $\eta=-\langle W\rangle/\langle Q_h\rangle \to \eta_{\rm C}=1-T_c/T_h$, we can verify that ${\rm var}(W_{\rm col})\propto C^{\rm col} \sim n^2$ (see Appendix \ref{App_sec_scaling}).  

A value of $\lambda_r > 1$ implies a lower relative noise-to-signal ratio for the collective quantum heat engine (QHE), as compared to the independent one, signifying higher consistency in the output of collective QHEs. The parameter $\lambda_r$ close to the Carnot bound is plotted in Fig.~\ref{fig:r_ratio_den}. The remarkable advantage offered by the collective effects is shown by large values of $\lambda_r$ in Fig. \ref{fig:r_ratio_den}, specially in the limit of large $n$ and $T_{\rm h}$. Clearly, high temperature $T_{\rm h}$  of the hot bath and large system size $n$ allow the operation of the collective QHE with high reliability in the work output. Interestingly, this regime is also the most beneficial for high mean work output $\langle W_{\rm col}\rangle \sim n^2$, as discussed in \cite{niedenzu18cooperative} and \cite{latune20collective}. 
Further,  $\lambda_r$ increases linearly with $n$ for $\beta_{\rm h} \to 0$, implying that the consistency of the collective engines increases extensively with system size. Remarkably, the quadratic advantage in reliability persists even for the more general Hamiltonians given in Eqs.~\eqref{eq:HLMG}, and \eqref{eq:Hint} with $x > 1$ [see Figs. \ref{fig:r_ratio_den}, \ref{fig:tur}, and Appendices \ref{App_sec_scaling} and \ref{App_sec_LMG} for details]. The collective advantage persists if the steady state is restricted to the $j=ns$ subspace, and is of the general form $\rho^{ss}(\beta)= \sum_{m=-ns}^{ns} \exp[{-\beta \epsilon_m}]/{Z}_{ns} |ns,m\rangle\langle ns,m|,\;{Z}_{ns}=\sum_{m=-ns}^{ns} \exp[{-\beta \epsilon_m}]$ where $\epsilon_m$ is the respective eigen-energy of the state $\ket{ns, m}$. Here the operator $\mathcal{A}$ introduced before can be identified as $\mathcal{A}=\sum_{m=-ns}^{ns} \epsilon_m/n \ket{ns,m}\bra{ns,m}$.  We summarize the above main results in Table \ref{table}.
\setlength{\tabcolsep}{2pt} 
\renewcommand{\arraystretch}{2} 
\begin{table}[h!]
\centering
\begin{tabular}{ |c|c|c| } 
\hline
 Quantity & Relation with heat capacity  & Scaling \\
\hline
Average work & $\langle W_{\rm col}\rangle \propto {C^{\rm col}(\theta_h)}/{\theta_h^2}$ & $n^2$~\cite{latune20collective}\\
 Work variance & $\text{var}(W_{\rm col}) \propto {C^{\rm col}(\theta_h)}/{\theta_h^2}$ &  $n^2$ \\ 
Reliability advantage & $\lambda_\text{r} \propto { C^{\rm col}(\theta_h)}/{C^{\rm ind}(\theta_h)}$ & $n$ \\
\hline
\end{tabular}
\caption{ Here we summarize the main results, which are valid in the limits $\eta\to\eta_{\rm C}$ and $T_h\to \infty$ with $\theta_h=\beta_h\omega_h$. The proofs are discussed in the main text and detailed in the Appendix.}
\label{table}
\end{table}

\section{ Entropy Production and TUR}
Having demonstrated the collective advantage in the reliability of a many-body correlated heat engine, we turn to the issue of entropy production. It is desirable to have an engine with low entropy production to enhance efficiency. However, it is known that the entropy production and the reliability are not independent of each other and there exists a lower bound for the product of entropy production and the inverse of the square of the reliability in the case of classical systems~\cite{pietzonka2018universal,horowitz2020thermodynamic}, which has been shown to hold true in many quantum systems as well~\cite{menczel2021thermodynamic,das2022precision}. For our collective engine, we can show that the enhanced reliability does not come at the cost of large entropy production; i.e., their product can be made smaller than the standard TUR bound.
The efficiency $\eta$ of an engine is related to the entropy production $\langle \Sigma\rangle$ through the relation: $\langle\Sigma\rangle = -\frac{\langle Q_{\rm h}\rangle}{T_{\rm h}} - \frac{\langle Q_{\rm c}\rangle}{T_{\rm c}} = \frac{\langle W \rangle}{T_{\rm c}}\left(\frac{\eta_{\rm C}}{\eta} - 1\right)$.

For the collective heat engine, the desired low fluctuating output work is obtained at a thermodynamic cost of an increased entropy production $\langle \Sigma\rangle$, following the inequality in the large $n$-limit \cite{sacchi2021multilevel}:
\begin{equation}
    \left(\frac{1}{r_{\rm col}}\right)^2 = \frac{f(\{\beta_i\omega_i\})}{\langle\Sigma_{\rm col}\rangle}-1 \;\geq\; \frac{2}{\langle\Sigma_{\rm col}\rangle} -1.
    \label{eq:tur_large_n}
\end{equation}
where $f(\{\beta_i\omega_i\})=(\beta_c \omega_c-\beta_h \omega_h) (a/b)\geq 2$ (see Appendix \ref{App_sec_tur}),
$a=\cosh (\beta_c \omega_c-\beta_h \omega_h)+\cosh \beta_c \omega_c+\cosh \beta_h \omega_h-3$ and
$b={\sinh (\beta_c \omega_c-\beta_h \omega_h)-\sinh \beta_c \omega_c+\sinh \beta_h \omega_h}$. Consequently, we can arrive at a  trade-off, quantified by the thermodynamic uncertainty $\mathcal{Q}=\langle\Sigma\rangle/r^2$~\cite{kalaee2021violating,horowitz2020thermodynamic}.
For a classical system $\mathcal{Q}\geq 2$, this is known as the standard TUR which provides a lower bound for $\mathcal{Q}$. For the $n$-qubit collective Otto engine, this is given by
\begin{equation} \mathcal{Q}_{\rm col} \geq 2-\langle\Sigma_{\rm col}\rangle, \label{eq:tur}\end{equation}
which further sharpens the TUR in the quantum case. Since the  entropy production $\langle\Sigma_{\rm col}\rangle\geq 0$, the collective engine TUR, Eq.~\eqref{eq:tur}, may violate the standard TUR bound, implying once again, a collective effect induced improvement in performance. Recently, such violations were observed for quantum-coherent and quantum-entangled systems~\cite{agarwalla2018assessing,ptaszynski2018coherence,guarnieri2019thermodynamics,cangemi2020violation,kalaee2021violating,sacchi2021multilevel}. 

In  Fig. \ref{fig:tur} the uncertainty $\mathcal{Q}$ as well as its lower bound are plotted for the collective and independent cases. The figure inset shows that in the large $n$ and $T_h$ limit, we have $\mathcal{Q}<2$; this is the region where the collective effects lead to a maximal advantage over the independent case as shown in Fig.~\ref{fig:r_ratio_den}. As in the case of reliability, the collective advantage in  $\mathcal{Q}$ extends to WM with inter-particle interactions, such as the LMG model [Eq. \eqref{eq:HLMG}], or with $x$-particle interactions [Eq. \eqref{eq:Hint}].

\section{Discussion}
In this manuscript, we considered QHEs modeled by generic many-body quantum systems collectively coupled to thermal baths. We showed that for a broad class of steady states arising due to collective system-bath coupling, we can get a quadratic advantage in reliability, as compared to their independent counterparts. We then explicitly showed this collective advantage in reliability for collective QHEs with non-interacting as well as interacting multi-spin WMs. Interestingly, this collective advantage increases with increasing temperature $T_h$ of the hot bath as well as size $n$ of the working medium. We also studied TUR in such collective QHEs; the advantage due to collective effects persists in this case as well, in the form of low thermodynamic uncertainty and TUR bound, as compared to the independent QHEs, for high  $T_h$. Furthermore, $\mathcal{Q}_{\rm col} < 2$ for high $T_h$ and large $n$, thereby violating the standard TUR bound in these regimes. Our analysis not only shows a way of realizing highly reliable many-body quantum machines but also extends the regime of validity of collective advantage to interacting models studied in many-body physics.

Several existing platforms can be expected to be ideal for experimentally realizing such collective QHEs. For example, experimental realization of collective effects is well established in cavity systems, such as by using Rydberg atoms in a cavity~\cite{PhysRevLett.49.117}; recently, the collective effect in the form of superabsorption in an organic microcavity was used to charge a quantum battery ~\cite{quach2022superabsorption}. The collective phenomenon of superradiance was observed in atomic sodium~\cite{gross1976observation} as well as in quantum dots~\cite{scheibner2007superradiance}. Recent advancements in techniques to assemble atoms with great control using optical tweezers~\cite{rui2020subradiant,de2019defect,glicenstein2020collective,barredo2016atom,endres2016atom}, and optical lattices~\cite{kumar2018sorting,greif2016site,sherson2010single,bakr2010probing} is also a promising development that can be used to design desired many-body quantum systems. 

Miniaturization of machines is one of the major aims of research in science and technology. However, this is associated with the challenge of fluctuations increasing with decreasing system size \cite{reif09}. In that context, the collective QHEs studied in this manuscript can be highly beneficial, thereby paving a path to realizing highly reliable mesoscopic quantum thermal machines.


\section*{ACKNOWLEDGEMENTS}

V.M. acknowledges support from Science and Engineering Research Board (SERB) through MATRICS (Project No.
MTR/2021/000055) and a Seed Grant from IISER Berhampur. S.V. acknowledges support from the Government of India DST-QUEST Grant No. DST/ICPS/QuST/Theme-4/2019.

\appendix

\section{ Work output of an Otto engine} \label{App_sec_work}
The total work output of an Otto engine is the sum of the work done on the system and by the system during the adiabatic unitary strokes. That is,
\begin{eqnarray}
\langle W\rangle &=& \langle W_{1}\rangle + \langle W_{2}\rangle\nonumber\\ 
&=&Tr\{ [H_h-H_c] \rho_c\} + Tr\{ [H_c-H_h)] \rho_h\} \nonumber \\
&=& (\omega_h-\omega_c) Tr\{\mathcal{G} [\rho_h-\rho_c]\}.
\end{eqnarray}
We have used the following definitions and identities here and in the following derivations:
$H_{\rm v} = \omega_{\rm v} \mathcal{G}$, $\rho_{\rm v} =\rho^{ss}(\theta_{\rm v})$, $\theta_{\rm v}=\beta_{\rm v}\omega_{\rm v},\;{\rm v}=\{c,h\}$,
$\Delta \eta = \eta_C-\eta = \frac{\theta_c-\theta_h}{\beta_c\omega_h}$,  $\eta_C$ is the Carnot efficiency, and 
$\Delta\theta = \theta_c-\theta_h = \beta_c\omega_h \Delta\eta$. The $\mathcal{G}$ could be $n (J_z/n)^x$ or $\left[ \frac{1}{n}(\mathcal{J}_x^2+\mathcal{J}_y^2)+\gamma \mathcal{J}_z \right]$. The collective angular momentum is defined as $\mathcal{J}_i := \sum_{k=1}^n J^k_{i}$, where $J_i^k$ is the spin operator for the $k$th spin, along the direction $i = x, y,z$. For $s=1/2$, $J_i^k =\sigma_i^k/2$ which is the Pauli matrix along $i$ direction times $1/2$.

Near the Carnot limit ($\Delta \eta \approx 0$), the work can be written as 
\begin{eqnarray}
\langle W \rangle &=& (\omega_c-\omega_h) \beta_c\omega_h \Delta\eta \; \frac{\partial}{\partial \theta}  Tr\{\mathcal{G} \rho(\theta)\}\arrowvert_{\theta_h} + O(\Delta \eta^2).\nonumber\\
\label{app_eqW}
\end{eqnarray}
 One can show that, 
 \ba
 \frac{\partial}{\partial \theta} Tr\{\mathcal{G} \rho(\theta)\}\arrowvert_{\theta_h} = -\frac{C(\theta_h)}{k_B \theta_h^2},
 \label{app_eqC}
 \ea
for both independent and collective cases, where $C$ is the heat capacity of the spin system. And also, $\omega_c-\omega_h=(\beta_c\omega_h -\beta_h\omega_h -\beta_c\omega_h \Delta \eta)/\beta_c $. Thus from Eqs. \eqref{app_eqW} and \eqref{app_eqC}, the average work is related to heat capacity as ~\cite{latune20collective},
\ba
\langle W\rangle =  -\Delta\eta \omega_h^2 (\beta_c-\beta_h)  \;   \frac{C(\theta_h)}{k_B \theta_h^2} + O(\Delta \eta^2).
\label{app_eqWC}
\ea

\begin{figure}[h]
         \centering
         \includegraphics[width=0.98\linewidth]{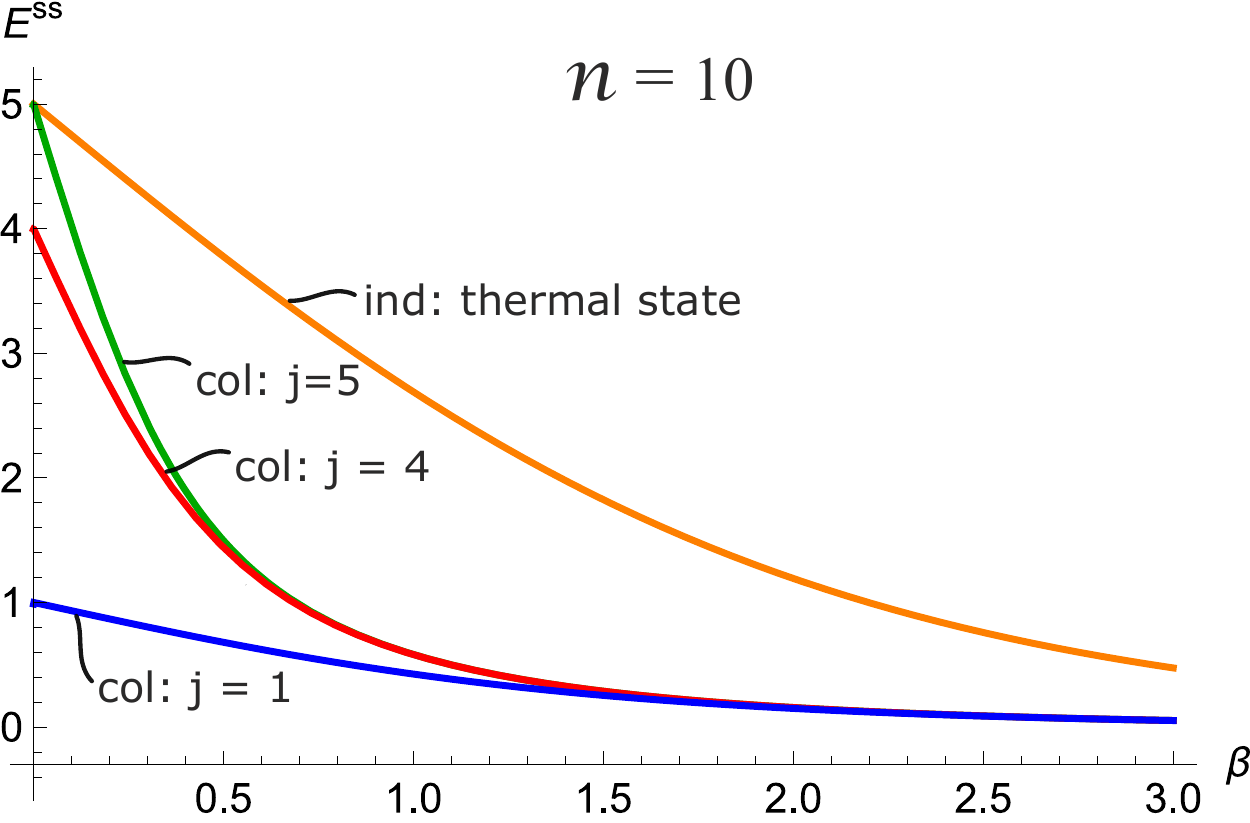}
      \caption{ The steady state energies are plotted as a function of the inverse temperature $\beta$ for independent and collective cases with $H=\omega \mathcal{J}_z$, $\hbar=\omega=1$, $s=1/2$ and $n=10$. For the collective case, the steady state is restricted to different $j$ subspaces and the curves are labeled accordingly; for the independent case, the steady state is given by the usual thermal state. Note that the heat capacity is  proportional to the slope of the steady-state energy. As shown in Table I of the main text, all quantities of interest can be expressed in terms of heat capacity. Therefore the different scaling advantages discussed in the main text as well as in this appendix can be understood from the slopes of the curves at large temperatures ($\beta\to 0$). The plot shows that for large $j$ values the collective heat capacities are larger than the independent case and it diminishes as $j$ decreases and below a certain value of $j$, its value becomes smaller than that of the independent case.}
        \label{app_fig:Ess}
\end{figure}

For an $n$ qubit system with $H = \omega \mathcal{J}_z$, inverse temperature $\beta$, and the steady state given in Eq.~(3) of the main text, the heat capacity becomes,
\begin{eqnarray}
C^{\rm col} &=& -k_B\beta^2\frac{\partial E^{ss}}{\partial \beta} \;=\; -k_B\beta^2\frac{\partial}{\partial \beta} Tr( H\rho^{ss})\nonumber\\
&=& \frac{1}{4} k_B {\beta}^2  {\omega}^2 \left[\text{csch}^2\frac{{\beta} {\omega}}{2}-(n+1)^2 \text{csch}^2\frac{ (n+1) {\beta\omega}}{2}\right],\nonumber\\
\end{eqnarray}
 where $E^{ss} =\frac{1}{2}n\hbar\omega + Tr( H\rho^{ss})$ is the steady state energy of the working medium (see Fig.~\ref{app_fig:Ess} and the reference \cite{latune2019thermodynamics}). At large temperatures, the above equation reduces to 
\ba \label{app_eq:Cns}
C^{\rm col}\approx \frac{1}{12} k_B \beta^2\omega^2 n(n+2),
\ea
which finally gives, $\langle W\rangle\propto n^2$. Similarly, it can be shown that for a steady state restricted to the spin $j$ subspace,  $C^{\rm col}/(\beta\omega)^2 \sim j(j+1)$ for $\beta \to 0$ (also see Fig.\ref{app_fig:Ess}). Equation~(\ref{app_eq:Cns}) is a special case of this with $j=n/2$.

\section{Characteristic function for a collective $n$ spin-$s$ system with $H=\omega \mathcal{J}_z$} \label{App_sec_nspin}

For an engine made of $n$ spin s systems with $H=\omega \mathcal{J}_z$, the characteristic function can be expressed as, $\chi = A/B$, where,
\begin{eqnarray*}
A &=& \left[e^{(2 n s+1)(\theta_h+i \gamma_1 \omega_h)}-e^{i (2 n s+1) (\gamma_1 \omega_c+\gamma_2 \omega_h)} \right]\nonumber\\
& & \; \left[ e^{i (2 n s+1) \gamma_1 \omega_h}-e^{(2 n s+1) [\theta_c+i (\gamma_1 \omega_c+\gamma_2 \omega_h)]} \right]\\
& & \;\; \sinh \frac{\theta_c}{2} \text{csch}\frac{(2ns+1)\theta_c}{2} \sinh \frac{\theta_h}{2} \text{csch}\frac{(2n s+1)\theta_h}{2} \nonumber\\
& & \quad e^{-n s \big\{\theta_c+\theta_h+2 i [\gamma_1 (\omega_c+\omega_h)+\gamma_2 \omega_h]\big\}}, \\
B &=& \left[e^{i \gamma_1 \omega_h}-e^{\theta_c+i (\gamma_1 \omega_c+\gamma_2 \omega_h)}\right] \nonumber\\
& & \; \left[e^{(\theta_h+i \gamma_1 \omega_h)}-e^{i (\gamma_1 \omega_c+\gamma_2 \omega_h)}\right],\;  \theta_{c(h)}=\beta_{c(h)}\omega_{c(h)}.
\end{eqnarray*}
For $s=1/2$ n-qubit engine, the general spin $s$ characteristic function simplifies to $\chi = A_1/B_1$, where,
\begin{eqnarray*} \label{eq:AB}
A_1 &=& \sinh \frac{\theta_c}{2} \sinh \frac{\theta_h}{2} \text{csch}\frac{ \theta_c (n+1)}{2} \text{csch}\frac{\theta_h (n+1)}{2} \\
& & \quad \left[e^{i \gamma_1 (n+1)\omega_h}-e^{(n+1) (\theta_c+i (\gamma_1\omega_c+\gamma_2\omega_h))}\right]\\
& & \qquad \left[e^{(n+1) (\theta_h+i \gamma_1 \omega_h )}-e^{i (n+1) (\gamma_1\omega_c+\gamma_2\omega_h)}\right] \\
& & \qquad\qquad e^{-\frac{n}{2} \big\{\theta_c+\theta_h +i 2[(\gamma_1+\gamma_2) \omega_h+ \gamma_1 \omega_c]\big\}},\\
B_1 &=& \left[e^{i \gamma_1\omega_h}-e^{\theta_c+i (\gamma_1\omega_c+\gamma_2\omega_h)}\right] \\ 
& & \qquad \left[e^{ (\theta_h+i \omega_h\gamma_1)}-e^{i (\gamma_1\omega_c+\gamma_2\omega_h)}\right].
\end{eqnarray*}

\section{ Work fluctuations} \label{App_sec_var}
The output work fluctuations of a frictionless Otto engine originate from the thermal energy fluctuations at the two thermalization strokes. Since the energy fluctuations are related to the heat capacity as $\sigma_E^2=C/(k_B \beta^2)$, one would expect the work fluctuations to be equal to the sum of the energy fluctuations at the two thermal strokes. We can show that this is indeed true for Hamiltonians,  such as the LMG model
\be \label{app_eq:HLMG}
H= \omega(t) \left[ \frac{1}{n}(\mathcal{J}_x^2+\mathcal{J}_y^2)+\gamma \mathcal{J}_z \right], 
\ee

or  with $x$-body inter-particle interactions, such as 
\ba
H=n\omega(t) (\mathcal{J}_z/n)^x~~~~\text{for~$x \geq 1$},
\label{app_eq:Hint}
\ea
if the density matrix is given by an appropriate Gibbs's distribution in $j=ns$ subspace, and the variance of the work takes the form, 
\ba
\text{var}(W)&=& (\omega_c-\omega_h)^2 \Big\{ \frac{C(\theta_h)}{k_B \theta_h^2} +  \frac{C(\theta_c)}{k_B \theta_c^2} \Big\} \label{app_eqvar0}\\
&\approx& \frac{2(\beta_c-\beta_h)^2}{k_B \beta_c^2\beta_h^2} C(\theta_h) +O(\Delta\eta)\nonumber\\
& =& 2\omega_h^2\eta_{\rm C}^2 \frac{C(\theta_h)}{k_B\theta_h^2} +O(\Delta\eta).
\label{app_eqvar}
\ea
The above equation \eqref{app_eqvar} holds for both the independent and collective coupling to the thermal baths. Again for the above cases, it can be shown that $C^{\rm col}(\theta_h)/\theta_h^2\propto n^2$. 
Thus the reliability for the collective bath coupling becomes $r_{\rm col}\propto n$. Further, $r_{\rm ind}\propto\sqrt{n}$ implies $\lambda_r = r_{\rm col}^2/r_{\rm ind}^2 \propto n$ which is a quadratic advantage for the collective case over the independent analog.
\section{ $n^2$ scaling of work, its variance and heat capacity for $H=n\omega {(\mathcal{J}_z/n)^x}$ } \label{App_sec_scaling}
Let the Hamiltonian be $H=n\omega {(\mathcal{J}_z/n)^x}$, where $\mathcal{J}_z=\frac{1}{2}\sum_i\sigma_z^i$. Following the steps given in the appendices \ref{App_sec_work} and \ref{App_sec_var}, we can show that 
\ba
\langle W\rangle =  -\Delta\eta \omega_h^2 (\beta_c-\beta_h)  \;   \frac{C(\theta_h)}{k_B \theta_h^2} + O(\Delta \eta^2) \nonumber
\ea
and
\ba
\text{var}(W) \;=\;  2\omega_h^2\eta_{\rm C}^2 \frac{C(\theta_h)}{k_B\theta_h^2} +O(\Delta\eta),\nonumber
\ea
for this case as well. Let the steady state be of the form $\rho^{ss}= e^{-\beta\omega H}/Z$ where $Z= \tr(e^{-\beta\omega H})$. Then heat capacity can be obtained using standard arguments as 
\be \label{app_eq:C_varH} C= -k_B \beta^2 \frac{\partial}{\partial \beta} Tr\left[ H \rho \right] = k_B \beta^2 {\rm var}(H).\ee

If the steady state is restricted to the $j=n/2$ irreducible subspace with Gibbs's distribution in the common eigenbasis of $J^2$ and $\mathcal{J}_z$, $\arrowvert j,m\rangle_i$ basis, then we get, \[\langle H\rangle=({n\omega/Z_{n/2}})\sum_m (m/n)^x e^{-\beta n \omega (m/n)^x},\]
and
\[\langle H^2\rangle =({n^2\omega^2/Z_{n/2}})\sum_m (m/n)^{2x} e^{-\beta n \omega (m/n)^{x}}.\] 
 The index $i\in [1; l_j]$ in $\arrowvert j,m\rangle_i$, where $l_j$ denotes the multiplicity of the eigenspaces associated with the eigenvalue of $\mathcal{J}^2$ operator. The $l_j$ can be explicitly expressed as~\cite{mandel1995optical}, \[l_j=\frac{(2j+1)\;n!}{(\frac{n}{2}+j+1)!\;(\frac{n}{2}-j)!},\]
which becomes $1$ for $j=n/2$. At large temperatures, that is inverse temperature $\beta\to 0$, the variance simplifies to 
\ba
\text{var}(H)&=& \langle H^2\rangle-\langle H\rangle^2 \nonumber\\
&=& \frac{n^2 \omega^2}{(n+1)}  \sum_{m} \left(\frac{m}{n}\right)^{2x} - \frac{n^2 \omega^2}{(n+1)^2} \left(\sum_{m} \left(\frac{m}{n}\right)^x\right)^2.\nonumber\\
\label{eqvargen}
\ea
For large $n$ values, the summations in Eq. \eqref{eqvargen} can be replaced with integration by setting $q=m/n$ giving, 
\ba\label{app_eq:varH_Jx}
\text{var}(H) \approx n^2 \omega^2 \left[\int_{-1/2}^{1/2} q^{2x} dq - \left(\int_{-1/2}^{1/2} q^{x} dq\right)^2\right].\nonumber\\
\ea
Here the $\text{var}(H)$ reduces to $n^2\frac{\omega^2 x^2}{4^x (2x+1)(x+1)^2}$ for even $x$, and $n^2\frac{\omega^2 }{4^x (2x+1)}$ for odd $x$. Thus both work and its variance are found to be proportional to $n^2$ for $x\geq 1$.

\section{ LMG system as the WM} 
\label{App_sec_LMG}

We consider a Lipkin-Meshkov-Glick (LMG) model given by the Hamiltonian,
\be 
H= \omega(t) \left[ \frac{1}{n}(\mathcal{J}_x^2+\mathcal{J}_y^2)+\gamma \mathcal{J}_z \right], 
\ee
as our working medium.
\setlength{\tabcolsep}{7.5pt} 
\renewcommand{\arraystretch}{3} 
\begin{center}
\begin{table}[h]
\begin{tabular}{ |c|c|c|c| } 
\hline
 Result & Scaling & Proof \\
\hline
$\langle W_{\rm col}\rangle \propto \frac{C^{\rm col}(\theta_h)}{\theta_h^2}$ & $n^2$ & Ref.~\cite{latune20collective}, Appendices~\ref{App_sec_work}, \\
 & & \ref{App_sec_scaling}, and \ref{App_sec_LMG} \\
$\text{var}(W_{\rm col}) \propto \frac{C^{\rm col}(\theta_h)}{\theta_h^2}$ &  $n^2$ &  Appendices~\ref{App_sec_var}, \ref{App_sec_scaling}, and \ref{App_sec_LMG}\\ 
 $r_{\rm col} \propto \sqrt{\frac{C^{\rm col}(\theta_h)}{\theta_h^2}}$ & $n$ &  Appendices~\ref{App_sec_scaling} and \ref{App_sec_LMG}\\
$\lambda_\text{r} \propto \frac{ C^{\rm col}(\theta_h)}{C^{\rm ind}(\theta_h)}$ & $n$ &   Appendices~\ref{App_sec_scaling} and \ref{App_sec_LMG}\\
\hline
\end{tabular}
\caption{The above main results are tabulated which are valid in the limits $\eta\to\eta_\text{C}$ and $T_h\to \infty$. We have defined $\theta_h=\beta_h\omega_h$.}
\label{app_table}
\end{table}
\end{center}
Following the same steps used to obtain the Eqs.~(\ref{app_eqWC}) and (\ref{app_eqvar}), we can show that near the Carnot efficiency, the work and its variance are proportional to the heat capacity. That is, $W\propto {C(\theta_h)}/{\theta_h^2}$ and $\text{var}(W)\propto {C(\theta_h)}/{\theta_h^2}$. Additionally, for a steady state of the form
\ba
\rho^{ss}(\beta\omega)&=&\sum_{m=-n/2}^{n/2} \frac{1}{Z_{\frac{n}{2}}}\; e^{-\beta \omega \left[\frac{1}{4n}n(n+2)-4m^2 + m\gamma\right]} \nonumber\\  & & \qquad\qquad\qquad\qquad \Ket{\frac{n}{2},m} \Bra{\frac{n}{2},m},\nonumber
\ea
we have $ \text{var}(H)\approx n^2\omega^2\left(\frac{1}{180}+\frac{\gamma^2}{12}\right)$. The partition function is given by $Z_{n/2}=\sum_{m=-n/2}^{n/2} e^{-\beta \omega \left[\frac{n(n+2)-4m^2}{4n} + m\gamma\right]}$. The above $\text{var}(H)$ is obtained for large $n$ values by replacing the summation with integration as we did in the Eq.~(\ref{app_eq:varH_Jx}). Thus from Eq.~(\ref{app_eq:C_varH}) we get the heat capacity $C\propto n^2$ giving an $n$ scaling for the reliability $r$. The main results of the above sections are tabulated in table~\ref{app_table}.

\section{ TUR for an $n$-qubit system with $H=\omega \mathcal{J}_z$} \label{App_sec_tur}
Now, using the moments we get the following inequality in the large $n$-limit:
\begin{equation}
    \left(\frac{1}{r_{\rm col}}\right)^2 = \frac{f(\{\beta_i\omega_i\})}{\langle\Sigma_{\rm col}\rangle}-1 \;\geq\; \frac{2}{\langle\Sigma_{\rm col}\rangle} -1.
    \label{app_eq:tur_large_n}
\end{equation}
where $f(\{\beta_i\omega_i\})=(\beta_c \omega_c-\beta_h \omega_h) (a/b)$, $a=\cosh (\beta_c \omega_c-\beta_h \omega_h) +\cosh \beta_c \omega_c +\cosh \beta_h \omega_h -3$ and
$b=\sinh (\beta_c \omega_c -\beta_h \omega_h)$ $-\sinh \beta_c \omega_c +\sinh \beta_h \omega_h$. Putting $\theta_c=\theta_h+\Delta$ in the function $f(\theta_c,\theta_h)=(\theta_c-\theta_h) (a/b)$ gives
\ba
f(\Delta,\theta_h) = \frac{\Delta  \left[3-\cosh \Delta -\cosh (\Delta +\theta_h)-\cosh \theta_h\right]}{\sinh \Delta -\sinh (\Delta +\theta_h)+\sinh \theta_h}.\nonumber\\
\ea
For large $\theta_h$ limit this becomes, $f(\Delta) = \Delta \coth \Delta/2 \geq 2$, and for other ranges of parameter values a simple numerical analysis yields minima as 2. We have used this result to obtain the inequality Eq.~\eqref{app_eq:tur_large_n}. Consequently, one can arrive at a  trade-off, quantified by the thermodynamic uncertainty $\mathcal{Q}=\langle\Sigma\rangle/r^2$. For the $n$-qubits collective Otto engine with $H=\omega(t)J_z$, trade-off can be found as 
\begin{equation} \mathcal{Q}_{\rm col}^{(1)}\geq 2-\langle \Sigma_{\rm col}^{(1)}\rangle, \label{app_eq:tur}\end{equation}
which further sharpens the TUR in the collective case. Here, $\langle \Sigma_{\rm col}^{(1)}\rangle$ and $\mathcal{Q}_{\rm col}^{(1)}$ are the collective entropy production and the quantity $\mathcal{Q}$ for $x=1$. 

 For an $n$-independent qubit engine, the uncertainty
\[\mathcal{Q}_{\rm ind}\;=\; \frac{\text{var}(W_{\rm ind}) \langle \Sigma_{\rm ind}\rangle}{\langle W_{\rm ind}\rangle^2} \;=\; \frac{\text{var}(W_{\rm ind}) \langle \Sigma_{\rm ind}\rangle}{\langle W_{\rm ind}\rangle^2} {\Big\arrowvert}_{n=1},\]
which reduces the lower bound on the thermodynamic uncertainty to a single qubit bound, that is, $\mathcal{Q}_{\rm ind}\geq 2-\langle \Sigma_{\rm ind}(n=1)\rangle$. Here, $\langle \Sigma_{\rm ind} (n=1)\rangle$ is the single spin engine entropy production. For brevity we may denote the $n$ spin entropy production $\langle \Sigma_{\rm ind} (n)\rangle$ with $\langle \Sigma_n\rangle$.   Also note that $\mathcal{Q}_{\rm ind}\geq 2-\langle \Sigma_1\rangle \geq 2-\langle \Sigma_n\rangle$, giving the single qubit bound as the tighter bound than the $2-\langle \Sigma_n\rangle$.


\bibliographystyle{ieeetr} 

\begin{thebibliography}{10}

\bibitem{kosloff14quantum}
Ronnie Kosloff and Amikam Levy.
\newblock Quantum heat engines and refrigerators: Continuous devices.
\newblock {\em Annual Review of Physical Chemistry}, 65(1):365--393, 2014.

\bibitem{klimovsky15thermodynamics}
David Gelbwaser-Klimovsky, Wolfgang Niedenzu, and Gershon Kurizki.
\newblock {Thermodynamics of quantum systems under dynamical control.}
 \newblock {\em Advances In Atomic, Molecular, and Optical Physics}
 \newblock { Vol.~64 (Elsevier, New York, 2015), Chap. 12, pp. 329–407.}

\bibitem{vinjanampathy16quantum}
Sai Vinjanampathy and Janet Anders.
\newblock Quantum thermodynamics.
\newblock {\em Contemporary Physics}, 57(4):545--579, 2016.

\bibitem{goold16the}
John Goold, Marcus Huber, Arnau Riera, L{\'{\i}}dia del Rio, and Paul Skrzypczyk.
\newblock The role of quantum information in thermodynamics
{\textemdash} a topical review.
\newblock {\em Journal of Physics A: Mathematical and Theoretical}, 49(14):143001, Feb 2016.

\bibitem{bhattacharjee21quantum}
Sourav Bhattacharjee and Amit Dutta.
\newblock Quantum thermal machines and batteries.
\newblock {\em The European Physical Journal B}, 94(12):239, Dec 2021.

\bibitem{mukherjee21many}
Victor Mukherjee and Uma Divakaran.
\newblock Many-body quantum thermal machines.
\newblock {\em Journal of Physics: Condensed Matter}, 33(45):454001, Aug 2021.

\bibitem{myers22quantum}
Nathan~M. Myers, Obinna Abah, and Sebastian Deffner.
\newblock Quantum thermodynamic devices: From theoretical proposals to experimental reality.
\newblock {\em AVS Quantum Science}, 4(2):027101, 2022.

\bibitem{PhysRevLett.125.240602}
Carollo Federico, Brandner Kay, and Lesanovsky Igor.
\newblock Nonequilibrium Many-Body Quantum Engine Driven by Time-Translation Symmetry Breaking.
\newblock {\em Physical Review Letters}, 125(24):240602, 2020.
 
\bibitem{PhysRevLett.127.190604}
Tajima Hiroyasu, and Funo Ken.
\newblock Superconducting-like Heat Current: Effective Cancellation of Current-Dissipation Trade-Off by Quantum Coherence.
\newblock {\em Physical Review Letters},127(19):190604, 2021.

\bibitem{PhysRevLett.86.5188}
Robert Raussendorf and Hans~J. Briegel.
\newblock A One-Way Quantum Computer.
\newblock {\em Phys. Rev. Lett.}, 86:5188, May 2001.

\bibitem{kielpinski2002architecture}
David Kielpinski, Chris Monroe, and David~J Wineland.
\newblock Architecture for a large-scale ion-trap quantum computer.
\newblock {\em Nature} (London), 417(6890):709--711, 2002.

\bibitem{monroe2014large}
C~Monroe, R~Raussendorf, A~Ruthven, KR~Brown, P~Maunz, L-M Duan, and J~Kim.
\newblock Large-scale modular quantum-computer architecture with atomic memory and photonic interconnects.
\newblock {\em Physical Review A}, 89(2):022317, 2014.

\bibitem{PhysRevD.23.1693} Caves, Carlton~M.
 \newblock {Quantum-mechanical noise in an interferometer}
 \newblock {\em Phys. Rev. D}, 23(8):1693--1708, 1981.
 
 \bibitem{PhysRevA.46.R6797}
D.~J. Wineland, J.~J. Bollinger, W.~M. Itano, F.~L. Moore, and D.~J. Heinzen.
\newblock Spin squeezing and reduced quantum noise in spectroscopy.
\newblock {\em Phys. Rev. A}, 46:R6797--R6800, Dec 1992.

\bibitem{giovannetti2004quantum}
Vittorio Giovannetti, Seth Lloyd, and Lorenzo Maccone.
\newblock Quantum-enhanced measurements: beating the standard quantum limit.
\newblock {\em Science}, 306(5700):1330--1336, 2004.

\bibitem{PhysRevA.33.4033}
Bernard Yurke, Samuel~L. McCall, and John~R. Klauder.
\newblock Su(2) and su(1,1) interferometers.
\newblock {\em Phys. Rev. A}, 33:4033, Jun 1986.

\bibitem{PhysRevA.57.4736}
Jonathan~P. Dowling.
\newblock Correlated input-port, matter-wave interferometer: Quantum-noise limits to the atom-laser gyroscope.
\newblock {\em Phys. Rev. A}, 57:4736--4746, Jun 1998.

\bibitem{PhysRevA.54.R4649}
J.~J~. Bollinger, Wayne~M. Itano, D.~J. Wineland, and D.~J. Heinzen.
\newblock Optimal frequency measurements with maximally correlated states.
\newblock {\em Phys. Rev. A}, 54:R4649--R4652, Dec 1996.

\bibitem{demkowiczchapter}
R~Demkowicz-Dobrzanski, M~Jarzyna, and J~Kolodynski.
\newblock Quantum limits in optical interferometry, 
\newblock {\em Progress in Optics}, Vol. 60 (Elsevier, New York, 2015). Chap. 4, pp. 345–435.

\bibitem{dowling2008quantum}
Jonathan~P Dowling.
\newblock Quantum optical metrology--the lowdown on high-n00n states.
\newblock {\em Contemporary physics}, 49(2):125--143, 2008.

\bibitem{jaseem2018quantum}
Noufal Jaseem, S~Omkar, and Anil Shaji.
\newblock Quantum critical environment assisted quantum magnetometer.
\newblock {\em Journal of Physics A: Mathematical and Theoretical}, 51(17):175309, 2018.

\bibitem{jaseem2017two}
Noufal Jaseem and Anil Shaji.
\newblock Two-mode gaussian product states in a lossy interferometer.
\newblock {\em Quantum Information Processing}, 16(9):1--14, 2017.

\bibitem{PhysRevA.82.042305}
Tatsuhiko Koike and Yosuke Okudaira.
\newblock Time complexity and gate complexity.
\newblock {\em Phys. Rev. A}, 82:042305, Oct 2010.

\bibitem{PhysRevA.82.042319}
Srinivas Sridharan, Mile Gu, Matthew~R. James, and William~M. McEneaney.
\newblock Reduced-complexity numerical method for optimal gate synthesis.
\newblock {\em Phys. Rev. A}, 82:042319, Oct 2010.

\bibitem{esposito09nonequilibrium}
Massimiliano Esposito, Upendra Harbola, and Shaul Mukamel.
\newblock Nonequilibrium fluctuations, fluctuation theorems, and counting statistics in quantum systems.
\newblock {\em Rev. Mod. Phys.}, 81:1665--1702, Dec 2009.

\bibitem{breuer02}
H.~P. Breuer and F.~Petruccione.
\newblock {\em The Theory of Open Quantum Systems}.
\newblock (Oxford University Press, Oxford, 2002).

\bibitem{mukherjee20anti}
Victor Mukherjee, Abraham~G. Kofman, and Gershon Kurizki.
\newblock Anti-zeno quantum advantage in fast-driven heat machines.
\newblock {\em Communications Physics}, 3(1):8, Jan 2020.

\bibitem{das20quantum}
Arpan Das and Victor Mukherjee.
\newblock Quantum-enhanced finite-time otto cycle.
\newblock {\em Phys. Rev. Research}, 2:033083, Jul 2020.

\bibitem{wiedmann21non}
Michael Wiedmann, J{\"u}rgen~T. Stockburger, and Joachim Ankerhold.
\newblock Non-markovian quantum otto refrigerator.
\newblock {\em The European Physical Journal Special Topics}, 230(4):851--857, Jun 2021.

\bibitem{modi}
Sai Vinjanampathy and Kavan Modi.
\newblock Correlations, operations and the second law of thermodynamics.
\newblock {\em International Journal of Quantum Information}, 14(06):1640033, 2016.

\bibitem{suri}
Nishchay Suri, Felix~C Binder, Bhaskaran Muralidharan, and Sai Vinjanampathy.
\newblock Speeding up thermalisation via open quantum system variational optimisation.
\newblock {\em The European Physical Journal Special Topics}, 227(3):203--216, 2018.

\bibitem{erdman19maximum}
P~A Erdman, V~Cavina, R~Fazio, F~Taddei, and V~Giovannetti.
\newblock Maximum power and corresponding efficiency for two-level heat engines and refrigerators: optimality of fast cycles.
\newblock {\em New Journal of Physics}, 21(10):103049, Oct 2019.

\bibitem{vasco21maximum}
Vasco Cavina, Paolo~A. Erdman, Paolo Abiuso, Leonardo Tolomeo, and Vittorio Giovannetti.
\newblock Maximum-power heat engines and refrigerators in the fast-driving regime.
\newblock {\em Phys. Rev. A}, 104:032226, Sep 2021.

\bibitem{erdman22identifying}
Paolo~A. Erdman and Frank No{\'e}.
\newblock Identifying optimal cycles in quantum thermal machines with reinforcement-learning.
\newblock {\em npj Quantum Information}, 8(1):1, Jan 2022.

\bibitem{khait22optimal}
Ilia Khait, Juan Carrasquilla, and Dvira Segal.
\newblock Optimal control of quantum thermal machines using machine learning.
\newblock {\em Phys. Rev. Research}, 4:L012029, Mar 2022.

\bibitem{rossnagel14nanoscale}
J.~Ro\ss{}nagel, O.~Abah, F.~Schmidt-Kaler, K.~Singer, and E.~Lutz.
\newblock Nanoscale heat engine beyond the carnot limit.
\newblock {\em Phys. Rev. Lett.}, 112:030602, Jan 2014.

\bibitem{niedenzu18quantum}
Wolfgang Niedenzu, Victor Mukherjee, Arnab Ghosh, Abraham~G. Kofman, and Gershon Kurizki.
\newblock Quantum engine efficiency bound beyond the second law of thermodynamics.
\newblock {\em Nature Communications}, 9(1):165, Jan 2018.

\bibitem{rossnage16a}
Johannes Ro{\ss}nagel, Samuel~T. Dawkins, Karl~N. Tolazzi, Obinna Abah, Eric Lutz, Ferdinand Schmidt-Kaler, and Kilian Singer.
\newblock A single-atom heat engine.
\newblock {\em Science}, 352(6283):325--329, 2016.

\bibitem{jan17squeezed}
Jan Klaers, Stefan Faelt, Atac Imamoglu, and Emre Togan.
\newblock Squeezed thermal reservoirs as a resource for a nanomechanical engine beyond the carnot limit.
\newblock {\em Phys. Rev. X}, 7:031044, Sep 2017.

\bibitem{mslennikov19quantum}
Gleb Maslennikov, Shiqian Ding, Roland Habl{\"u}tzel, Jaren Gan, Alexandre Roulet, Stefan Nimmrichter, Jibo Dai, Valerio Scarani, and Dzmitry Matsukevich.
\newblock Quantum absorption refrigerator with trapped ions.
\newblock {\em Nature Communications}, 10(1):202, Jan 2019.

\bibitem{klatzow19experimental}
James Klatzow, Jonas~N. Becker, Patrick~M. Ledingham, Christian Weinzetl, Krzysztof~T. Kaczmarek, Dylan~J. Saunders, Joshua Nunn, Ian~A. Walmsley, Raam Uzdin, and Eilon Poem.
\newblock Experimental demonstration of quantum effects in the operation of microscopic heat engines.
\newblock {\em Phys. Rev. Lett.}, 122:110601, Mar 2019.

\bibitem{pal20experimental}
Soham Pal, Sushant Saryal, Dvira Segal, T.~S. Mahesh, and Bijay~Kumar Agarwalla.
\newblock Experimental study of the thermodynamic uncertainty relation.
\newblock {\em Phys. Rev. Research}, 2:022044(R), May 2020.

\bibitem{campisi16the}
Michele Campisi and Rosario Fazio.
\newblock The power of a critical heat engine.
\newblock {\em Nature Communications}, 7(1):11895, Jun 2016.

\bibitem{revathy20universal}
Revathy B.~S, Victor Mukherjee, Uma Divakaran, and Adolfo del Campo.
\newblock Universal finite-time thermodynamics of many-body quantum machines from kibble-zurek scaling.
\newblock {\em Phys. Rev. Research}, 2:043247, Nov 2020.

\bibitem{fogarty20a}
Thom{\'{a}}s Fogarty and Thomas Busch.
\newblock A many-body heat engine at criticality.
\newblock {\em Quantum Science and Technology}, 6(1):015003, Nov 2020.

\bibitem{halpern19quantum}
Nicole Yunger~Halpern, Christopher~David White, Sarang Gopalakrishnan, and Gil Refael.
\newblock Quantum engine based on many-body localization.
\newblock {\em Phys. Rev. B}, 99:024203, Jan 2019.

\bibitem{chiaracane20quasiperiodic}
Cecilia Chiaracane, Mark~T. Mitchison, Archak Purkayastha, G\'eraldine Haack, and John Goold.
\newblock Quasiperiodic quantum heat engines with a mobility edge.
\newblock {\em Phys. Rev. Research}, 2:013093, Jan 2020.

\bibitem{niedenzu18cooperative}
Wolfgang Niedenzu and Gershon Kurizki.
\newblock Cooperative many-body enhancement of quantum thermal machine power.
\newblock {\em New Journal of Physics}, 20(11):113038, Nov 2018.

\bibitem{carlos}
Carlos~A Perez-Delgado and Sai Vinjanampathy.
\newblock Coherent parallelization of universal classical computation.
\newblock {\em New Journal of Physics}, 23(12):123015, 2021.

\bibitem{angsar19collectively}
Angsar Manatuly, Wolfgang Niedenzu, Ricardo Rom\'an-Ancheyta, Bar\ifmmode \imath \else \i \fi{}\ifmmode \mbox{\c{s}}\else~\c{s}\fi{} \ifmmode~\mbox{\c{C}}\else \c{C}\fi{}akmak, \"Ozg\"ur~E. M\"ustecapl\ifmmode \imath \else \i \fi{}o\ifmmode~\breve{g}\else \u{g}\fi{}lu, and Gershon Kurizki.
\newblock Collectively enhanced thermalization via multiqubit collisions.
\newblock {\em Phys. Rev. E}, 99:042145, Apr 2019.

\bibitem{latune20collective}
C~L Latune, I~Sinayskiy, and F~Petruccione.
\newblock Collective heat capacity for quantum thermometry and quantum engine enhancements.
\newblock {\em New Journal of Physics}, 22(8):083049, Aug 2020.

\bibitem{watanabe20quantum}
Gentaro Watanabe, B.~Prasanna Venkatesh, Peter Talkner, Myung-Joong Hwang, and Adolfo del Campo.
\newblock Quantum statistical enhancement of the collective performance of multiple bosonic engines.
\newblock {\em Phys. Rev. Lett.}, 124:210603, May 2020.

\bibitem{seeding}
Michal Hajdu{\v{s}}ek, Parvinder Solanki, Rosario Fazio, and Sai Vinjanampathy.
\newblock Seeding crystallization in time.
\newblock {\em Physical Review Letters}, 128(8):080603, 2022.

\bibitem{souza22collective}
Leonardo da~Silva Souza, Gonzalo Manzano, Rosario Fazio, and Fernando Iemini.
\newblock Collective effects on the performance and stability of quantum heat engines.
\newblock {\em Phys. Rev. E}, 106:014143, Jul 2022.

\bibitem{kamimura22quantum}
Shunsuke Kamimura, Hideaki Hakoshima, Yuichiro Matsuzaki, Kyo Yoshida, and Yasuhiro Tokura.
\newblock Quantum-enhanced heat engine based on superabsorption.
\newblock {\em Phys. Rev. Lett.}, 128:180602, May 2022.

\bibitem{klimovsky19cooperative}
David Gelbwaser-Klimovsky, Wassilij Kopylov, and Gernot Schaller.
\newblock Cooperative efficiency boost for quantum heat engines.
\newblock {\em Phys. Rev. A}, 99:022129, Feb 2019.

\bibitem{kloc19collective}
Michal Kloc, Pavel Cejnar, and Gernot Schaller.
\newblock Collective performance of a finite-time quantum otto cycle.
\newblock {\em Phys. Rev. E}, 100:042126, Oct 2019.

\bibitem{reif09} F.~Reif.
\newblock {\em Fundamentals of Statistical and Thermal Physics}.
\newblock {(Waveland, Long Grove, IL, 2009).}

\bibitem{juzar2021monitoring}
Jeongrak Son, Peter Talkner, and Juzar Thingna.
\newblock Monitoring quantum otto engines.
\newblock {\em PRX Quantum}, 2(4):040328, 2021.

\bibitem{sacchi2021multilevel}
Massimiliano~F Sacchi.
\newblock Multilevel quantum thermodynamic swap engines.
\newblock {\em Physical Review A}, 104(1):012217, 2021.

\bibitem{Sacchi2021boson}
Massimiliano~F. Sacchi.
\newblock Thermodynamic uncertainty relations for bosonic otto engines.
\newblock {\em Phys. Rev. E}, 103:012111, Jan 2021.

\bibitem{agarwalla2018assessing}
Bijay~Kumar Agarwalla and Dvira Segal.
\newblock Assessing the validity of the thermodynamic uncertainty relation in quantum systems.
\newblock {\em Physical Review B}, 98(15):155438, 2018.

\bibitem{ptaszynski2018coherence}
Krzysztof Ptaszy{\'n}ski.
\newblock Coherence-enhanced constancy of a quantum thermoelectric generator.
\newblock {\em Physical Review B}, 98(8):085425, 2018.

\bibitem{guarnieri2019thermodynamics}
Giacomo Guarnieri, Gabriel~T Landi, Stephen~R Clark, and John Goold.
\newblock Thermodynamics of precision in quantum nonequilibrium steady states.
\newblock {\em Physical Review Research}, 1(3):033021, 2019.

\bibitem{cangemi2020violation}
Loris~Maria Cangemi, Vittorio Cataudella, Giuliano Benenti, Maura Sassetti, and Giulio De~Filippis.
\newblock Violation of thermodynamics uncertainty relations in a periodically driven work-to-work converter from weak to strong dissipation.
\newblock {\em Physical Review B}, 102(16):165418, 2020.

\bibitem{kalaee2021violating}
Alex Arash~Sand Kalaee, Andreas Wacker, and Patrick~P Potts.
\newblock Violating the thermodynamic uncertainty relation in the three-level maser.
\newblock {\em Physical Review E}, 104(1):L012103, 2021.

\bibitem{ito2019universal}
Kosuke Ito, Chao Jiang, and Gentaro Watanabe.
\newblock Universal bounds for fluctuations in small heat engines.
\newblock {\em arXiv:1910.08096}.

\bibitem{horowitz2020thermodynamic}
Jordan~M Horowitz and Todd~R Gingrich.
\newblock Thermodynamic uncertainty relations constrain non-equilibrium fluctuations.
\newblock {\em Nature Physics}, 16(1):15--20, 2020.



\bibitem{Caneva_LMG}
Caneva Tommaso, Fazio Rosario, and Santoro Giuseppe~E.
  \newblock{Adiabatic quantum dynamics of the Lipkin-Meshkov-Glick model.}
  \newblock{ \em Physical Review B}, 78(10):104426, 2008.
  
\bibitem{larson2010circuit}
Larson Jonas.
\newblock{Circuit QED scheme for the realization of the Lipkin-Meshkov-Glick model.}
\newblock{\em {EPL (Europhysics Letters)}}, 90(5):54001, 2010.

\bibitem{PhysRevC.104.024305}
Cervia Michael~J., Balantekin A.B., Coppersmith~S.N., Johnson Calvin~W., Love Peter J., Poole~C., Robbins~K., and Saffman~M.
\newblock {Lipkin model on a quantum computer.}
\newblock {\em Phys. Rev. C}, 104(2):024305, 2021.

 \bibitem{lipkin1965validity}
 H.J.~Lipkin , N~Meshkov, and A.J~Glick.
\newblock Validity of many-body approximation methods for a solvable model:(I). Exact solutions and perturbation theory.
\newblock {\em Nuclear Physics}, 62(2):188--198, 1965.

\bibitem{latune2019thermodynamics}
Camille~L Latune, Ilya Sinayskiy, and Francesco Petruccione.
\newblock {Thermodynamics from indistinguishability: Mitigating and amplifying the effects of the bath}
\newblock {\em Phys. Rev. Research}, 1:033192, 2019.

\bibitem{mandel1995optical}
 Mandel Leonard, and Wolf Emil.
 \newblock {\em Optical coherence and quantum optics}.
 \newblock {(Cambridge university press, Cambridge, England, 1995).}

 \bibitem{PhysRevE.103.L060103} Saryal Sushant and Agarwalla Bijay Kumar.
 \newblock  Bounds on fluctuations for finite-time quantum Otto cycle.
 \newblock {\em Physical Review E}, 103(6):L060103, 2021.

\bibitem{denzler2020thesis}
Tobias Denzler.
\newblock {\em Fluctuations and correlations of quantum heat engines}.
\newblock Ph.D. thesis, University of Stuttgart, 2020.
 
\bibitem{pietzonka2018universal} Pietzonka Patrick and Seifert Udo.
\newblock {Universal trade-off between power, efficiency, and constancy in steady-state heat engines}
\newblock {\em Physical Review Letters}, 120(19):190602, 2018.

\bibitem{menczel2021thermodynamic} Menczel Paul, Loisa Eetu, Brandner Kay, and Flindt Christian.
\newblock {Thermodynamic uncertainty relations for coherently driven open quantum systems.}
\newblock {\em Journal of Physics A: Mathematical and Theoretical}, 54(31):314002, 2021.

\bibitem{das2022precision} Das Arpan, Shishira Mahunta, Agarwalla Bijay Kumar, and Mukherjee Victor.
\newblock {Precision bound in periodically modulated continuous quantum thermal machines.}
\newblock{ \em arXiv:2204.14005}.

\bibitem{PhysRevLett.49.117}
J.~M. Raimond, P.~Goy, M.~Gross, C.~Fabre, and S.~Haroche.
\newblock Collective absorption of blackbody radiation by rydberg atoms in a cavity: An experiment on bose statistics and brownian motion.
\newblock {\em Physical Review Letters}, 49:117, Jul 1982.

\bibitem{quach2022superabsorption}
James~Q Quach, Kirsty~E McGhee, Lucia Ganzer, Dominic~M Rouse, Brendon~W Lovett, Erik~M Gauger, Jonathan Keeling, Giulio Cerullo, David~G Lidzey, and Tersilla Virgili.
\newblock Superabsorption in an organic microcavity: Toward a quantum battery.
\newblock {\em Science advances}, 8(2):eabk3160, 2022.

\bibitem{gross1976observation}
M~Gross, C~Fabre, P~Pillet, and S~Haroche.
\newblock Observation of near-infrared dicke superradiance on cascading transitions in atomic sodium.
\newblock {\em Physical Review Letters}, 36(17):1035, 1976.

\bibitem{scheibner2007superradiance}
Michael Scheibner, Thomas Schmidt, Lukas Worschech, Alfred Forchel, Gerd Bacher, Thorsten Passow, and Detlef Hommel.
\newblock Superradiance of quantum dots.
\newblock {\em Nature Physics}, 3(2):106--110, 2007.

\bibitem{rui2020subradiant}
Jun Rui, David Wei, Antonio Rubio-Abadal, Simon Hollerith, Johannes Zeiher, Dan~M Stamper-Kurn, Christian Gross, and Immanuel Bloch.
\newblock A subradiant optical mirror formed by a single structured atomic layer.
\newblock { Nature (London)}, 583(7816):369--374, 2020.

\bibitem{de2019defect}
Daniel~Ohl De~Mello, Dominik Sch{\"a}ffner, Jan Werkmann, Tilman Preuschoff, Lars Kohfahl, Malte Schlosser, and Gerhard Birkl.
\newblock Defect-free assembly of 2D clusters of more than 100 single-atom quantum systems.
\newblock {\em Physical review letters}, 122(20):203601, 2019.

\bibitem{glicenstein2020collective}
Antoine Glicenstein, Giovanni Ferioli, Nikola {\v{S}}ibali{\'c}, Ludovic Brossard, Igor Ferrier-Barbut, and Antoine Browaeys.
\newblock Collective shift in resonant light scattering by a one-dimensional atomic chain.
\newblock {\em Physical Review Letters}, 124(25):253602, 2020.

\bibitem{barredo2016atom}
Daniel Barredo, Sylvain De~L{\'e}s{\'e}leuc, Vincent Lienhard, Thierry Lahaye, and Antoine Browaeys.
\newblock An atom-by-atom assembler of defect-free arbitrary two-dimensional atomic arrays.
\newblock {\em Science}, 354(6315):1021--1023, 2016.

\bibitem{endres2016atom}
Manuel Endres, Hannes Bernien, Alexander Keesling, Harry Levine, Eric~R Anschuetz, Alexandre Krajenbrink, Crystal Senko, Vladan Vuletic, Markus Greiner, and Mikhail~D Lukin.
\newblock Atom-by-atom assembly of defect-free one-dimensional cold atom arrays.
\newblock {\em Science}, 354(6315):1024--1027, 2016.

\bibitem{kumar2018sorting}
Aishwarya Kumar, Tsung-Yao Wu, Felipe Giraldo, and David~S Weiss.
\newblock Sorting ultracold atoms in a three-dimensional optical lattice in a realization of maxwell’s demon.
\newblock { Nature (London)}, 561(7721):83--87, 2018.

\bibitem{greif2016site}
Daniel Greif, Maxwell~F Parsons, Anton Mazurenko, Christie~S Chiu, Sebastian Blatt, Florian Huber, Geoffrey Ji, and Markus Greiner.
\newblock Site-resolved imaging of a fermionic mott insulator.
\newblock {\em Science}, 351(6276):953--957, 2016.

\bibitem{sherson2010single}
Jacob~F Sherson, Christof Weitenberg, Manuel Endres, Marc Cheneau, Immanuel Bloch, and Stefan Kuhr.
\newblock Single-atom-resolved fluorescence imaging of an atomic mott insulator.
\newblock { Nature (London)}, 467(7311):68--72, 2010.

\bibitem{bakr2010probing}
Waseem~S Bakr, Amy Peng, M~Eric Tai, Ruichao Ma, Jonathan Simon, Jonathon~I Gillen, Simon Foelling, Lode Pollet, and Markus Greiner.
\newblock Probing the superfluid--to--mott insulator transition at the single-atom level.
\newblock {\em Science}, 329(5991):547--550, 2010.

\end{thebibliography}

\end{document}